\documentclass[doublecol]{epl2} 
\usepackage{graphicx}
\usepackage{bm}
\usepackage{amsmath}
\usepackage{amsfonts}

\def\ra{\rangle}
\def\la{\langle}
\def\ua{\uparrow}
\def\da{\downarrow}

\def\sfB{\mathsf{B}}
\def\sfT{\mathsf{T}}
\def\sfTL{\mathsf{T}_\mathrm{L}}
\def\LG{L(G)}
\def\VL{V_\mathrm{L}}
\def\EL{E_\mathrm{L}}
\def\HS{H_\mathrm{S}}
\def\Hb{H_\mathrm{b}}
\def\Hf{H_\mathrm{f}}
\newcommand{\hal}[1]{\relax}
\title{Ferromagnetism in the Hubbard model with topological/non-topological flat bands}
\shorttitle{Title} 

\author{Hosho Katsura\inst{1,2} \and Isao Maruyama\inst{3} \and Akinori Tanaka\inst{4} \and Hal Tasaki\inst{5}}
\shortauthor{F. Author \etal}

\institute{                    
  \inst{1} Kavli Institute for Theoretical Physics, University of California - Santa Barbara, CA 93106, USA\\
  \inst{2} Cross-Correlated Materials Research Group (CMRG), ASI, RIKEN 
- Wako, Saitama 351-0198, Japan\\
  \inst{3} Graduate School of Engineering Science, Osaka University
- Toyonaka, Osaka 560-8531, Japan\\
  \inst{4} Department of General Education, Ariake National College of Technology - Omuta, Fukuoka 836-8585, Japan\\
  \inst{5} Department of Physics, Gakushuin University - Mejiro, Toshima-ku, Tokyo 171-8588, Japan}


\abstract{
We introduce and study two classes of Hubbard models with
magnetic flux or with spin-orbit coupling,
which have a flat lowest band separated from other bands by a nonzero gap.
We study the Chern number of the flat bands, and find that it is 
zero for the first class but can be
nontrivial in the second.
We also prove that the introduction of  on-site Coulomb repulsion leads to
ferromagnetism in both the classes.}

\begin{document}

\maketitle

\section{Introduction}
Motivated by the recent discovery of quantum spin Hall effect in band insulators with strong spin-orbit coupling (SOC)~\cite{Kane_Mele_PRL05, Bernevig_Science06, Koenig_Science07}, the topological classification of non-interacting electron systems has attracted a renewed interest~\cite{Qi, Ryu, Kitaev}. 
The states of the systems are characterized by the topological numbers linked to the presence or absence of gapless edge modes carrying electronic or spin current. 
In  integer quantum Hall systems~\cite{Prange_Girvin}, 
the first Chern number is directly connected to the quantized Hall conductance~\cite{TKNN},
which is one of the most famous examples of time reversal breaking insulators.
On the other hand, in the recently found time reversal invariant insulators,  the states are classified by the $\mathbb{Z}_2$ topological number.
Since a topological number remains invariant as long as the energy gap 
does not collapse, 
adiabatic transformation from the original model to a flat-band model, where all the bands are dispersionless, provides a useful tool for the classification~\cite{Qi, Ryu, Kitaev}.


The flat-band models also play an important role in a completely different context, i.e., rigorous examples of  ferro- or ferrimagnetism in the Hubbard model~\cite{Lieb, Mielke, Tasaki, Gulacsi}. 
In the models proposed by Mielke~\cite{Mielke} and by Tasaki~\cite{Tasaki}, the on-site Coulomb interaction leads to ferromagnetic ground states when the lowest flat band is half-filled. 
A common feature of these models is that the electron hoppings are frustrated and the lowest band is spanned by moderately localized eigenstates. 
Recently, similar localized states have been found in highly frustrated quantum magnets in strong magnetic fields~\cite{Schulenburg, Tsunetsugu} and optical lattice models~\cite{C_Wu}, and offer a playground for studying nonperturbative aspects of strongly correlated systems. 


These two subjects have  developed separately, and  
the effect of electron correlation in topological insulators 
has not been studied intensively.
\hal{This has been further modified.}
Here we study the Hubbard model with flat bands in the presence of magnetic flux or SOC, which bridges the two subjects.
In order to define the topological number, the gap between the flat band and  other bands is required. 
The possibility of a gapped flat band is already a nontrivial issue since in many cases the band touching occurs at some points in momentum space~\cite{Bergman} and a uniform magnetic field destroys the flatness
~\cite{Aoki_PRB_96}. 
In this Letter, we propose two classes of  tight binding models (TBM) which have a flat band separated from  other bands by a nonzero gap~\cite{Green}. 
The first class is a TBM on a line graph (e.g. checkerboard lattice) 
with non-uniform flux. 
For models in this class, we find that the flat band is always non-topological, i.e., the Chern number is always zero.  
We can study the effect of interaction rigorously and show that the ground states are ferromagnetic when the lowest band is half-filled for both magnetic-flux and SOC cases. 
The second class is a TBM embedded on a thin torus with a magnetic field perpendicular to the plane. 
Surprisingly, all the bands become flat if a special condition is satisfied. 
We calculate the topological numbers 
and show that the topological flat band indeed exists. 
We also study the effect of electron correlation and rigorously show that 
the ground states are ferromagnetic when the lowest Landau level (LLL) is
half-filled, i.e., $\nu=1/$odd, without using the LLL projection~\cite{MacDonald_PRL_96}.

\hal{This part has been considerably modified.}
\section{Non-topological flat band}
\begin{figure}[tb]
\begin{center}
\includegraphics[width=\columnwidth]{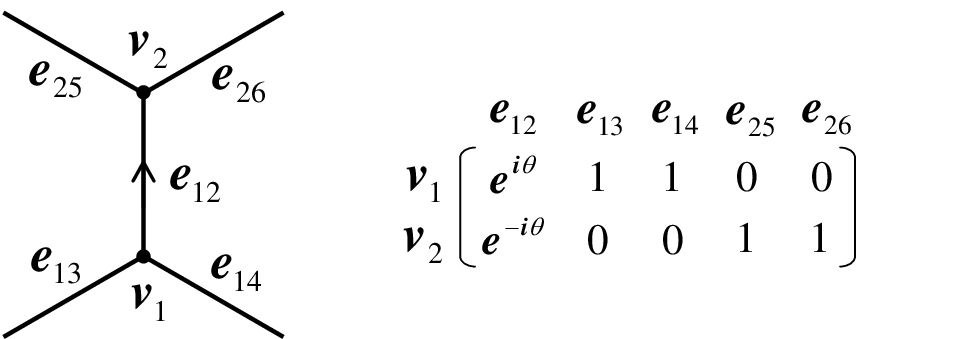}
\caption{A graph consisting of 2 vertices and 5 edges, and corresponding incidence matrix. The arrow indicates the sign of the phase factor $e^{i\theta}$.
}
\label{incidence}
\end{center}
\end{figure}

We start from the first class.
Let $G=(V, E)$ be a graph (lattice), where $V$ is the set of vertices (sites) and $E$  the set of edges (bonds). 
We assume that $G$ is twofold connected~\cite{twofold_graph}.
We define the incidence matrix $\sfB=(B_{ve})_{v \in V, e \in E}$ by assigning 
a nonzero complex number to $B_{ve}$ when the vertex $v$ is incident to the edge $e$,
and by setting $B_{ve}=0$ otherwise (see Fig.~\ref{incidence} for example).
\begin{figure}[tb]
\begin{center}
\includegraphics[width=\columnwidth]{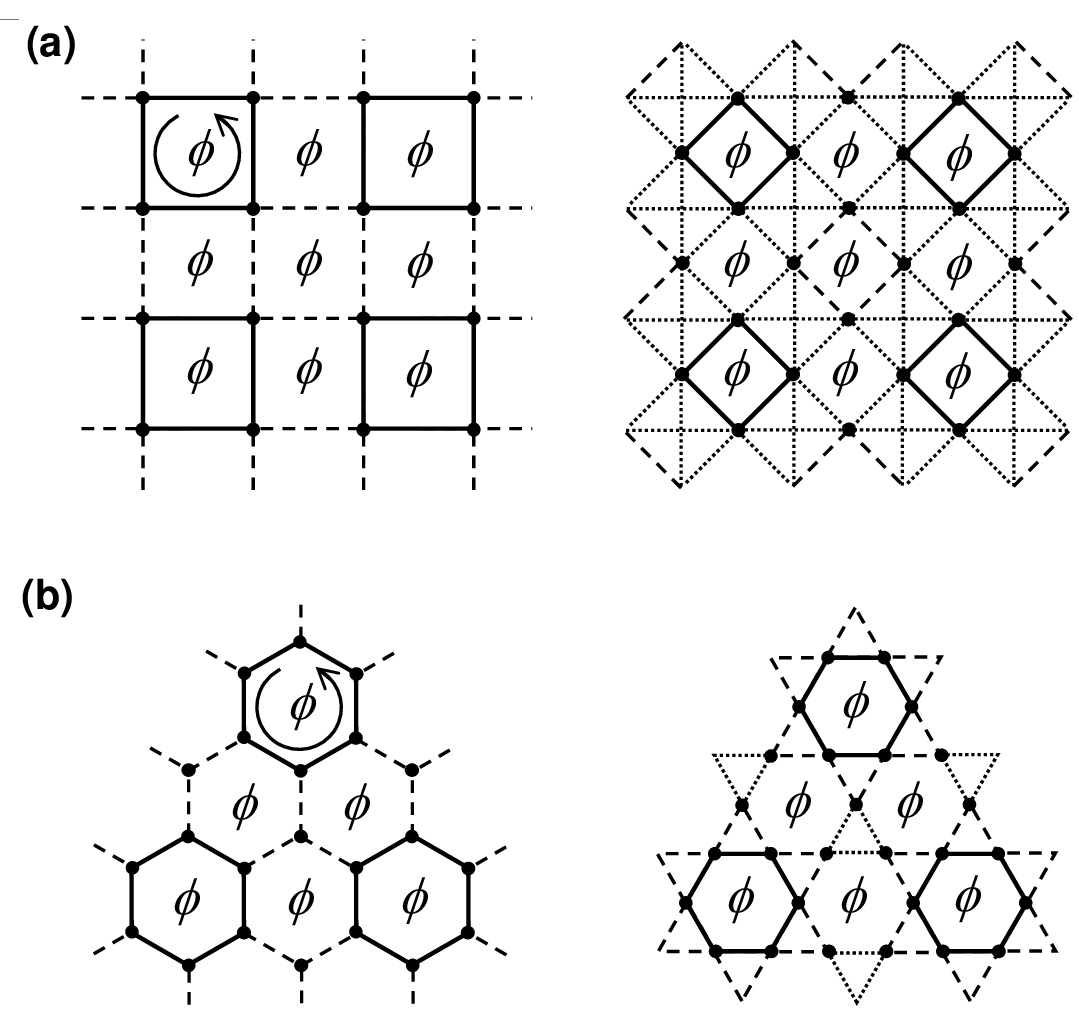}
\caption{
(a) Left: Tight-binding model on a square lattice with magnetic flux $\phi$ through each plaquette. Solid and broken edges belong to $E'$ and $E''$, respectively. 
The rotating arrow indicates the orientation of vertices. 
Right: Line graph of the left graph. 
The flux $\phi$ goes through diamond-shaped plaquettes.
The hopping amplitudes $|(\sfTL(x))_{ee'}|$ of solid, dotted, and broken edges are $1$, $\sqrt{x}$, and $x$ ($0 \le x \le 1$), respectively.
\hal{Some changes in the above.
I would denote a site of $\LG$ as $e,e'$.}
(b) Honeycomb lattice and its line graph (Kagom\'e lattice). 
Notations are the same as in (a).
}
\label{incidence1}
\end{center}
\end{figure}
Define the  line graph $\LG=(\VL, \EL)$ of $G$ as usual, by
first regarding (the midpoint of) each edge in $E$ as a vertex in $\VL$ (thus $\VL=E$),
and then connecting any pair of  vertices in $\VL$ (by an edge in $\EL$) when the corresponding pair of edges in $E$ share a common vertex. 
Fig.~\ref{incidence1}~(a) shows the square lattice and its line graph, the checkerboard lattice (ignore the differences in the bonds for the moment).
Let us consider TBMs on $G$ and $\LG$ with hopping matrices (single-particle Hamiltonians) $\sfT=\sfB\sfB^\dagger$ and $\sfTL=\sfB^\dagger\sfB$, respectively.
It is easily shown that 
\hal{I added the nonnegativity, and changed the order}
(T1)~all the eigenvalues of  $\sfT$ and $\sfTL$ are nonnegative, 
(T2)~nonzero eigenvalues of $\sfT$ and $\sfTL$ are 
identical, and 
(T3)~$\sfTL$ has at least $(|E|-|V|)$ zero-energy eigenstates~\cite{properties_of_T}. 
(T1) follows if one notes that both $\sfT$ and $\sfTL$ are positive
semidefinite.
(T3) is an immediate consequence of the fact that $\sfB$ is a $|V| \times |E|$ matrix.
To see (T2), let $\boldsymbol{\varphi}$ be an eigenvector 
of $\sfT$ with a nonzero eigenvalue $a$,
i.e, $\sfB\sfB^\dagger\boldsymbol{\varphi}=a\boldsymbol{\varphi}$.
Multiplying $\sfB^\dagger$ from the left, one finds
$\sfTL\tilde{\boldsymbol{\varphi}}=a\tilde{\boldsymbol{\varphi}}$
with a nonzero vector  $\tilde{\boldsymbol{\varphi}}=\sfB^\dagger\boldsymbol{\varphi}$.
To complete the proof we only need to 
repeat the same argument with $\sfT$ and $\sfTL$ switched~\cite{SUSY}.
For periodic systems, these zero energy eigenstates of $\sfTL$ form the lowest flat band.

Let us now show that the flat band in the TBM 
on $\LG$
may be gapped but is not topological. 
To be concrete we let $G$ be the square lattice~\cite{line_graph_gene}, and consider 
models with  a constant flux $\phi$ per plaquette.
For each edge $\la vv' \ra\in E$, define $\phi_{vv'}=-\phi_{v'v}\in\mathbb{R}$
so that $\phi_{v_1v_2}+\phi_{v_2v_3}+\phi_{v_3v_4}+\phi_{v_4v_1}=\phi$ (mod 1)
for any plaquette $\la v_1v_2v_3v_4 \ra$ oriented in the counterclockwise
direction.
\hal{I deleted the range of $\phi$, and declared that the above is a mod 1 relation.}
By setting $B_{ve}=\exp[\pi i\phi_{vv'}]$ if $e=\la vv' \ra$, we get a TBM
on $G$ with a uniform flux, and the corresponding TBM on the line graph $\LG$
with a uniform flux through the diamond plaquettes (see Fig.~\ref{incidence}~(b), still
ignoring the differences in the bonds)~\cite{cal_of_Tvv'}. 

It is useful for our proof  to introduce interpolating TBMs with an extra
parameter $0\le x\le 1$.
Consider a disjoint decomposition $E=E'\cup E''$  
as in Fig. \ref{incidence}(b), and redefine $B_{ve}$ as
$\sqrt{x}\exp[\pi i\phi_{vv'}]$ only if $e=\la vv' \ra\in E''$.
We denote by $\sfT(x)$ and $\sfTL(x)$ the corresponding 
hopping matrices.
Note that $\sfT(1)$ and $\sfTL(1)$ are the same as the original
$\sfT$ and $\sfTL$, respectively, and both
$\sfT(0)$ and $\sfTL(0)$ describe TBMs which decouple into
local pieces.
From the definition, one finds for any $0\le x\le1$ that
$\sfT(x)\ge\sfT(0)\ge\epsilon(\phi)$ where 
$\epsilon(\phi)$
is the lowest eigenvalue of $T(0)$~\cite{proof_of_1st_ineq}.
Suppose that $\phi$ is nonintegral.
Since an explicit calculation shows $\epsilon(\phi)>0$~\cite{lowest_energy}, 
the above (T2) implies that $\sfTL(x)$ has a nonzero gap
above the zero eigenvalue.
Recalling that the lowest flat band of $\sfTL$ is gapless for $\phi=0$,
we see that the above gap in $\sfTL$ originates from the flux.
\hal{SOme changes}


Now we investigate the Chern number defined by imposing twisted boundary conditions as in~\cite{Fukui_Hatsugai, Niu_Thouless_Wu} of the flat band
in the TBM on the line graph $\LG$.
Suppose that $\phi$ is nonintegral.
Since the Chern number is invariant as long as the energy gap does not close, 
one can evaluate it in the TBM with $\sfTL(x)$ for any $x$.
But since a decoupled system is insensitive to a twist in the boundary conditions,
the Chern number of the flat band in $\sfTL(0)$ is clearly zero.
This proves that the Chern number of the flat band in the original model with $\sfTL$ is zero.
\hal{Corrections have been made.}


Although the flat band is non-topological in this class of models, we can rigorously study the effect of electron correlation. 
We present two specific examples in the following.\\
\section{i) Kagom\'{e} ladder}
\begin{figure}[tb]
\begin{center}
\includegraphics[width=.95\columnwidth]{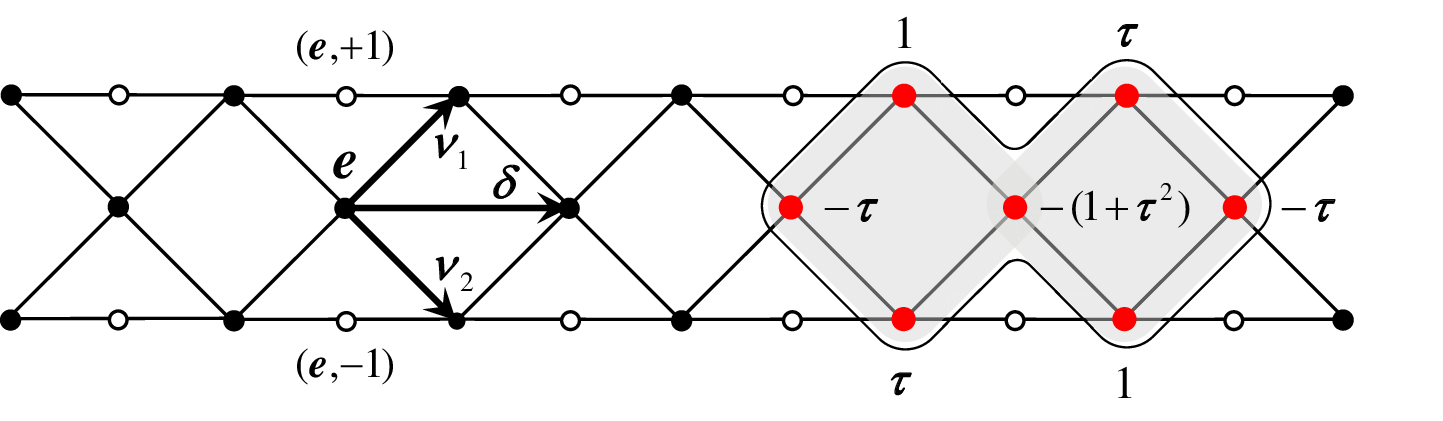}
\caption{A portion of the Kagom\'e ladder. 
Unfilled circles are vertices of the square ladder. 
The shaded region depicts the localized state $f^\dagger$ and
its amplitudes are indicated with $\tau=e^{i\theta/2}$. 
Fluxes through plaquette and triangle are $-\theta/(2\pi)$ and $0$, respectively.}
\label{kagome_ladder}
\end{center}
\end{figure}

The first example is the Hubbard model on the Kagom\'e ladder, the line graph of the square ladder, shown in Fig. \ref{kagome_ladder}. 
The sites in the mid-chain are labeled by $\bm{e}=n\bm{\delta}$ with $n=0, 1, \dots, 2L-1$ as shown in Fig.~\ref{kagome_ladder}. 
We denote vertices of the square ladder by $(\bm{e},l)$ with $l=\pm1$.
\hal{I want to define $B$ in the following.}
We let the incidence matrix element $B_{(\bm{e},l)\,\bm{e}^\prime}$ be
 $1$ if  $\bm{e}^\prime=\bm{e}$ or $\bm{e}-l\bm{\nu}_2$, $e^{i\theta/2}$ 
if $\bm{e}^\prime=\bm{e}+l\bm{\nu}_1$, and 0 otherwise.
Here $\theta$ is the parameter determining the flux, and 
$\bm{\nu}_1$ and $\bm{\nu}_2$ are the vectors indicated in Fig.~\ref{kagome_ladder}. 
The tight-binding Hamiltonian of our model is  then given by
\hal{I have deleted the range of $\theta$.}
\begin{equation}
 H_\mathrm{KL}  = 
 \mathop{\sum_{\sigma=\ua,\da}}_{\bm{e},\bm{e}'\in\VL}
 (\sfTL)_{\bm{e}\bm{e}'}
 c^\dagger_{\bm{e},\sigma}c_{\bm{e}',\sigma}=
 \mathop{\mathop{\sum_{\sigma=\ua,\da}}_{\bm{e}\in\VL^\mathrm{mid}}}_{l=\pm1} 
                   a_{(\bm{e},l),\sigma}^\dagger 
                   a_{(\bm{e},l),\sigma}
 \label{eq:KLHamiltonian}
\end{equation}
\hal{I explicitly used $\sfTL$ matrix.
I know that this trick of sum is ugly...
BTW I  realized that we had not defined the fermion operators, but that should be allowed.}%
where $\bm{e}$ in the right-hand side is summed over the $2L$ sites in the mid-chain,
and the $a$-operators are defined as 
\begin{equation}
 a_{(\bm{e},l),\sigma}=\sum_{\bm{e}^\prime} B_{(\bm{e},l)\,\bm{e}^\prime} c_{\bm{e}^\prime,\sigma}
                 = 
                   c_{\bm{e},\sigma}
                   + e^{i\theta/2}c_{\bm{e}+l\bm{\nu}_1,\sigma}
		   +c_{\bm{e}-l\bm{\nu}_2,\sigma}.
\end{equation}
We impose periodic boundary conditions in the chain direction
(see Fig. \ref{kagome_ladder}).
By a straightforward calculation, one finds that
\begin{eqnarray}
 f_{n,\sigma}^\dagger=-(e^{i\theta}+1)c^\dagger_{\bm{e},\sigma}
~~~~~~~~~~~~~~~~~~~~~~~~~~~~~~~~~~~~~~~~~ \nonumber \\      
+ \sum_{l=\pm1}(
                                    e^{i\theta/2}c^\dagger_{\bm{e}+l\bm{\nu}_1,\sigma} 
                                    +c^\dagger_{\bm{e}+l\bm{\nu}_2,\sigma}
                                    -e^{i\theta/2}c^\dagger_{\bm{e}+l\bm{\delta},\sigma}
                               )
\end{eqnarray} 
anticommutes with the $a$-operators. 
Therefore, the single-electron zero-energy states
are given by
$
 f_{n,\sigma}^\dagger|\Phi_0\ra
$,
where $|\Phi_0\rangle$ is the vacuum state.
The collection of these states forms a complete basis for the flat band.
\hal{I added the definition of $\Phi_0$ since I had deleted it.}
\hal{I did not understand what the sentence
``It is noted that the flat band is intact for any $\theta$.'' mean.}

We shall consider the case where the flat band is half-filled.  
In the non-interacting case, the $2L$-electron ground states are
highly degenerate and exhibit paramagnetism.
Let us add the Hamiltonian the standard 
on-site Coulomb repulsion 
\begin{equation}
H_U=U\sum_{\bm{e} \in \VL}n_{\bm{e},\uparrow}n_{\bm{e},\downarrow}, 
\end{equation}
where $n_{\bm{e},\sigma} = c^\dagger_{\bm{e},\sigma} c_{\bm{e},\sigma}$ is the number operator at a site $\bm{e}$ of the Kagom\'e ladder. 
Then the degeneracy is lifted and  
only the ferromagnetic states remain as the ground states,
which are
the zero energy eigenstate of both $H_\mathrm{KL}$ and $H_U$.
\hal{Separated the long sentence into two, and also included the content
of the endnote.}
To prove this claim, we introduce 
new fermion operators 
$
 d_{n,\sigma}^\dagger=e^{i\theta/2} f_{n,\sigma}^\dagger-f_{n+(-1)^n,\sigma}^\dagger,
$
and follow the standard strategy \cite{flatbandferro}.
\hal{I found this part very hard to understand.
I just wanted to say that the technique itself is standard, and
can be learned from the literature.}
The states $d^\dagger_{n,\sigma}|\Phi_0\ra$ 
form another basis for the flat band. 
By representing a ground state $|\Psi\ra$ 
in terms of the $d$-operators, 
we can firstly show   
that the zero-energy conditions for the on-site repulsion, 
$
 c_{\bm{e}^\prime,\da}c_{\bm{e}^\prime,\ua}|\Psi\ra=0
$, with sites 
$
 \bm{e}^\prime=n\bm{\delta}+(-1)^n\bm{\nu}_1
$
forbid the double occupancy of $d$-states. 
Then the same conditions with sites in the mid-chain  
imply that the ground states must be  
\hal{``conclude'' is changed to ``imply''}
$
 \left(\prod^{2L-1}_{n=0} d_{n,\ua}^\dagger\right)|\Phi_0\ra
$ 
and its SU(2) rotations. 
We note that even when the flat band is less than half-filled, ferromagnetic states are ground states but are not unique.  

It is also possible to consider the non-topological flat band model
associated with SOC. 
For the Kagom\'e ladder, the corresponding tight-binding Hamiltonian $H_\mathrm{KL}^\mathrm{SOC}$ is 
produced by replacing $e^{i\theta/2}$ in the definition of $a$-operators with $e^{i\sigma\theta/2}$ in Eq. (\ref{eq:KLHamiltonian}). 
This model has spin-dependent complex hoppings. 
Since for $\theta=\pi$, the model is reduced to the case of 
spin-independent hopping 
by a local gauge transformation,
we restrict ourselves to $\theta \ne \pi$.  
The single-electron zero-energy states of 
$H_\mathrm{KL}^\mathrm{SOC}$
are given by 
$\tilde{d}_{n,\sigma}^{\dagger}|\Phi_0\ra$,  
where 
$\tilde{d}_{n,\sigma}^\dagger$ is defined
as $d_{n,\sigma}^\dagger$ with 
\begin{eqnarray}
 \tilde{f}_{n,\sigma}^\dagger=-(e^{i\sigma\theta}+1)c^\dagger_{\bm{e},\sigma}
~~~~~~~~~~~~~~~~~~~~~~~~~~~~~~~~~~~~~~~~ \nonumber \\
+\sum_{l=\pm1}(
                 e^{i\sigma\theta/2}c^\dagger_{\bm{e}+l\bm{\nu}_1,\sigma} 
                 +c^\dagger_{\bm{e}+l\bm{\nu}_2,\sigma}
                 -e^{i\sigma\theta/2}c^\dagger_{\bm{e}+l\bm{\delta},\sigma})
                 \end{eqnarray}
in place of $f_{n,\sigma}^\dagger$.
In the $2L$-electron case, the ground states of the Hubbard Hamiltonian $H=H_\mathrm{KL}^\mathrm{SOC}+H_U$
are given by $\left(\prod^{2L-1}_{n=0}{\tilde{d}_{n,\sigma}^{\dagger}}\right)|\Phi_0\ra$ with $\sigma=\ua,\da$. 
In contrast to the spin-independent case, the SU(2) spin degeneracy is lifted and there are only two ground states.\\
\section{ii) Two-dimensional checkerboard lattice}
\begin{figure}[tb]
\begin{center}
\includegraphics[width=.7\columnwidth]{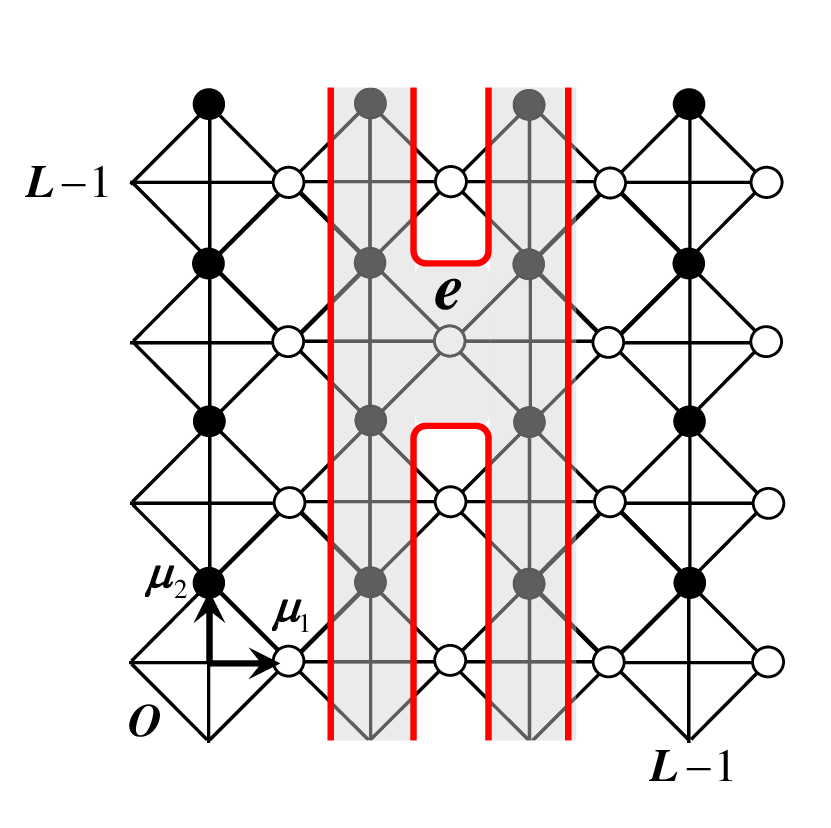}
\caption{
Checkerboard lattice (Line graph of the square lattice). 
Filled (unfilled) circles represent sites corresponding to the edges 
in $E_2$ ($E_1$). The shaded region depicts the localized state $d_{{\bm e}, \sigma}$ in Eq. (\ref{eq: checkerboard d}).
}
\label{checkerboard}
\end{center}
\end{figure}

The second is the Hubbard model on the checkerboard lattice,
the line graph of the square lattice whose vertex set is 
given by $V=[0,L-1]^2\cap\Bbb{Z}^2$ with an odd positive integer $L$.
The periodic boundary conditions are imposed in both directions. 
We label an element of the edge set $E$ of the square lattice
by its mid-point position. 
Let $\bm{\mu}_1=(1/2,0)$ and $\bm{\mu}_2=(0,1/2)$. 
For later use 
we decompose $E$ as $E=E_1\cup E_2$, 
where $E_l=\{\bm{e}=\bm{v}+\bm{\mu}_l|\bm{v}\in V\}$. 
Let the incidence matrix element $B_{\bm{v}\bm{e}}$ be 1 
if $\bm{e}=\bm{v}\pm \bm{\mu}_2$ or $\bm{e}=\bm{v}- \bm{\mu}_1$, 
$e^{i2(\bm{v}\bullet \bm{\mu}_2)\theta}$ if $\bm{e}=\bm{v}+ \bm{\mu}_1$,
and 0 otherwise.
Here $\theta$ is again the parameter determining the flux $\phi=-\theta/(2\pi)$.  
The tight-binding Hamiltonian of this model
is then given by
\begin{equation}
 H_\mathrm{CL}  = 
 \mathop{\sum_{\sigma=\ua,\da}}_{\bm{e},\bm{e}'\in\VL}
 (\sfTL)_{\bm{e}\bm{e}'}
 c^\dagger_{\bm{e},\sigma}c_{\bm{e}',\sigma}=
  \mathop{\sum_{\sigma=\ua,\da}}_{\bm{v}\in V} 
                   a_{\bm{v},\sigma}^\dagger 
                   a_{\bm{v},\sigma}
 \label{eq:sqHamiltonian}
\end{equation}
where the $a$-operators are defined as
\begin{align}
a_{\bm{v},\sigma} &=\sum_{\bm{e}} B_{\bm{v}\bm{e}}c_{\bm{e},\sigma}
 \notag \\
                  &=
                  e^{i2(\bm{v}\bullet
		  \bm{\mu}_2)\theta}c_{\bm{v}+\bm{\mu}_1,\sigma}
                   +c_{\bm{v}-\bm{\mu}_1,\sigma}
                   +c_{\bm{v}+\bm{\mu}_2,\sigma}
                   +c_{\bm{v}-\bm{\mu}_2,\sigma}.    
\end{align}
For each $\bm{e}\in E_1$ we define 
\begin{eqnarray}
d_{\bm{e},\sigma}^\dagger&=&
                 2c_{\bm{e},\sigma}^\dagger
                 - e^{i(2\bm{e}\bullet\bm{\mu}_2)\theta}
                  \sum_{l=0}^{L-1}(-1)^l c_{\bm{e}-\bm{\mu}-2l\bm{\mu}_2,\sigma}^\dagger
 \nonumber\\ 
		 &&- 
                     \sum_{l=0}^{L-1}(-1)^l
		     c_{\bm{e}+\bm{\mu}+2l\bm{\mu}_2,\sigma}^\dagger
\label{eq: checkerboard d}		     
\end{eqnarray}
where $\bm{\mu}=\bm{\mu}_1+\bm{\mu}_2$. 
These states are extended in one direction and localized in perpendicular one 
as shown in Fig. \ref{checkerboard}. 
It is easy to see that the $d$-operators are
anticommute with the $a$-operators, and
therefore the single-electron zero-energy states of $H_\mathrm{CL}$
are given by
$
 d_{\bm{e},\sigma}^\dagger|\Phi_0\ra
$.
The collection of these states forms 
a complete basis for the flat band.

As in the case of the Kagom\'e ladder, 
the fully polarized states, 
$
 \left(\prod_{\bm{e}\in E_1} d_{\bm{e},\ua}^\dagger\right)|\Phi_0\ra,
$ 
and its SU(2) rotations are the unique ground states of 
the Hubbard Hamiltonian $H=H_\mathrm{CL}+H_U$
when the electron number is $L^2$, i.e., the flat band is half-filled. 
Note that these ground states are simultaneous eigenstates of 
both $H_\mathrm{CL}$ and $H_U$ with zero-energy. 
This claim can be proved by following the same strategy.
Representing a ground state $|\Psi\ra$ 
in terms of the $d$-operators and 
noting that $\bm{e}$ in $E_1$ supports only $d_{\bm{e},\sigma}$,  
we can firstly show   
that the zero-energy conditions for the on-site repulsion, 
$
 c_{\bm{e},\da}c_{\bm{e},\ua}|\Psi\ra=0
$, with sites 
$
 \bm{e}\in E_1
$
forbid the double occupancy of $d$-states. 
Then, examing the same conditions with sites $\bm{e}\in E_2$
we 
arrive at the conclusion.


\section{Topological flat band}
\begin{figure}[tb]
\begin{center}
\includegraphics[width=\columnwidth]{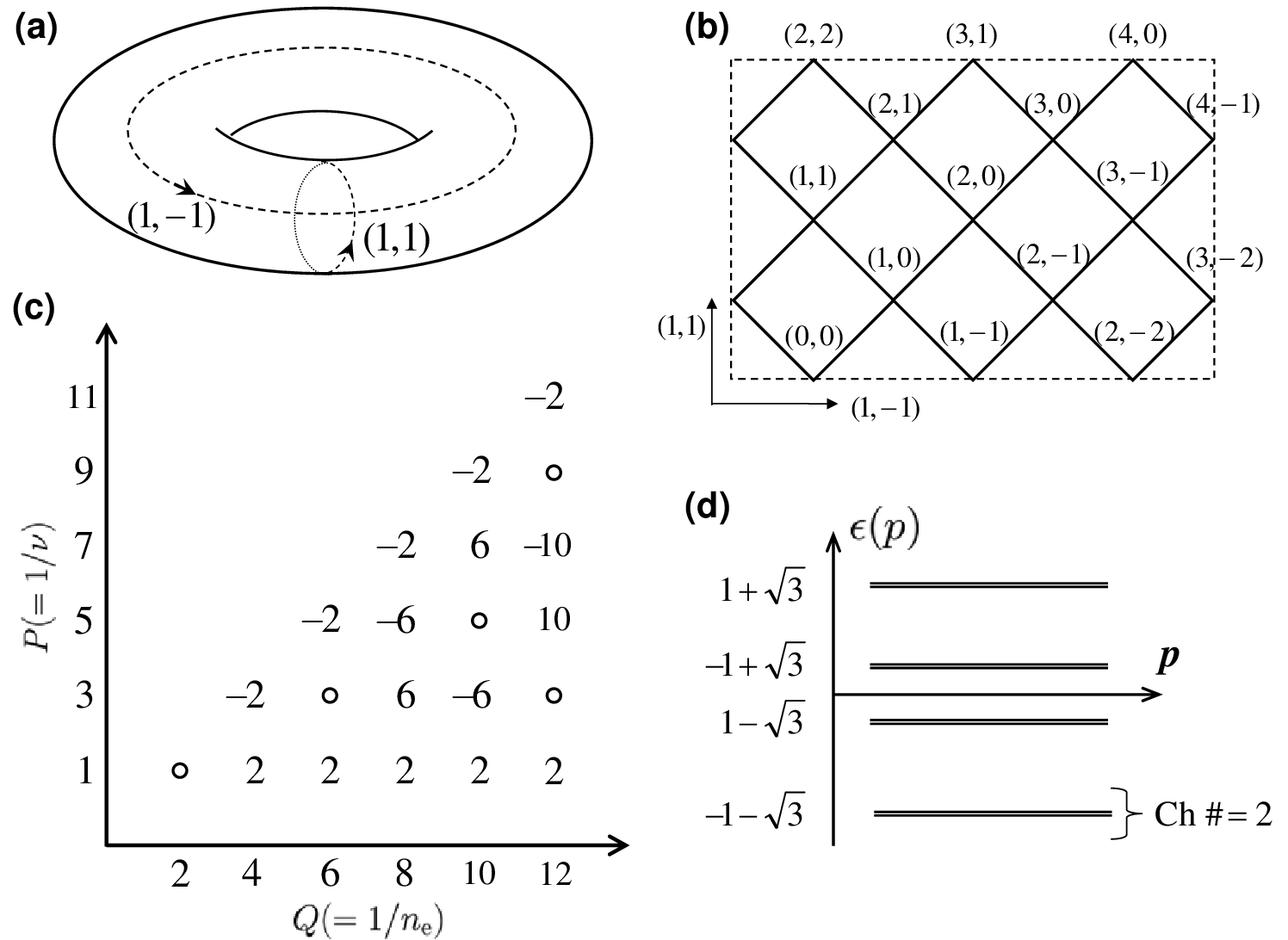}
\caption{a) Thin torus. Independent cycles in $(1,1)$ and $(1,-1)$ directions are indicated by the broken lines. b) Lattice structure for $Q=2$ and $L=3$. $(m,n)$ denotes the label for the vertex. Periodic boundary is indicated by the dashed line. 
c) Table for the sum of the Chern numbers of the lowest flat bands. $n_{\rm e}$ and $\nu$ are the electron density and the filling factor, respectively. 
The unfilled circles indicate that the Chern number is undefined since 
there is a gap closing caused by a twist in the boundary conditions. 
(d) Single-electron band structure for $P=1$ and $Q=4$. 
Each band is doubly degenerate (per spin). The sum of the Chern numbers of the lowest bands is $2$.}
\label{thin_torus}
\end{center}
\end{figure}

We turn to the second class.
\hal{I have made it compact.}
We consider the square lattice TBM embedded on a torus 
with a magnetic field perpendicular to the plane. Such a problem is known as the Hofstadter problem~\cite{Hofstadter} and has been extensively studied~\cite{Wiegmann,Hatsugai, Abanov}.
As we will show, all the bands become flat if the flux per plaquette and the number of sites along $(1,1)$ direction satisfy certain conditions.  
We shall use a notation as close to those in \cite{Hatsugai} as possible. 
The tight-binding Hamiltonian is given by $H_{\rm hop}=T_x+T_y+T^\dagger_x+T^\dagger_y$ with
\begin{eqnarray}
T_x &=& \sum_{\sigma=\ua,\da}\sum_{m,n}e^{i\theta^x_{m,n}}c^\dagger_{(m+1,n),\sigma}c_{(m,n),\sigma}, \\
T_y &=& \sum_{\sigma=\ua,\da}\sum_{m,n}e^{i\theta^y_{m,n}}c^\dagger_{(m,n+1),\sigma}c_{(m,n),\sigma},
\end{eqnarray}
where 
$(m,n)$ denote the vertices of the square lattice, and $\theta^x_{m,n}=(m+n)\pi \phi$ and $\theta^y_{m,n}=-(m+n+1)\pi \phi$. 
\hal{I thought $(m+n)\pi \phi$ is easier to read than $\pi \phi(m+n)$.}
The flux per plaquette is $\phi=P/Q$ with mutually prime $P$ and $Q$. 
Periodic boundary conditions are imposed both in $(1,1)$ and $(1,-1)$ directions. 
\hal{both ``in'' ... directions.}
From the Bloch theorem, we can assume the single-electron state to be of the form
\hal{I don't think we need a semicolon here.}
\begin{eqnarray}
|\Phi_\sigma(p_+,p)\ra &=& \sum_{m,n} \Psi_{m,n}(p_+,p) c^\dagger_{(m,n),\sigma}|\Phi_0\ra, \label{Bloch_wf} \\
\Psi_{m,n}(p_+,p) &=& e^{ip_+(m+n)+ip(m-n)}\psi_{m+n}(p_+,p),
\end{eqnarray}
where $\psi_{k+2Q}(p_+,p)=\psi_k(p_+,p)$, ($k=0, 1, ..., 2Q-1$). 
We now consider the thin torus case where the system size is finite along one of the cycles of the torus (see Fig. \ref{thin_torus}(a)). 
Suppose that the periodic boundary conditions in $(1,1)$ and $(1,-1)$ directions are given by
\begin{eqnarray} 
\Psi_{m+Q,n+Q}(p_+,p) &=& \Psi_{m,n}(p_+,p) \\
\Psi_{m+L,n-L}(p_+,p) &=& \Psi_{m,n}(p_+,p). 
\end{eqnarray}
Then they yield $p_+=\pi l/Q$ (mod $\pi/Q$) with $l\in \mathbb{Z}$, and $p=k\pi/L$ 
with $k=0, 1, ..., L-1$.
Therefore, if $Q$ is even, $p_+=\pi/2$ (mod $\pi/Q$) and it automatically satisfies the {\it mid-band condition}. 
This condition is closely related to the symmetry described by the quantum group $U_q(sl_2)$~\cite{Wiegmann, Hatsugai}. 

Henceforth we shall focus on the case of even $Q$ and show that all the bands are flat, namely, the single-electron energy $\epsilon(p)$ is independent of $p$. The Schr\"odinger equation for $|\Phi_\sigma(\pi/2,p)\ra$ is written as
\begin{equation}
i(q^{l+1}-q^{-(l+1)}) u_{l+1}+i(q^l-q^{-l}) u_{l-1}=\epsilon(p)u_l,
\label{midband_eq}
\end{equation} 
where $q=e^{i\pi P/Q}$, and $u_l$ are defined through the unitary transformation
\begin{equation}
\psi_k(\pi/2, p)=\frac{1}{\sqrt{2Q}}\sum^{2Q-1}_{l=0}q^{lk}e^{-ilp} u_l.
\end{equation} 
%
It is now obvious that $\epsilon(p)$ does not depend on $p$ because there is no $p$-dependence in the LHS of Eq. (\ref{midband_eq}). 
The single-electron energy for $P=1$, $Q=4$ is shown in Fig. \ref{thin_torus}~(d). 
Each band is doubly degenerate for the reason discussed below. 
Interestingly the  lowest flat band in this construction can be topological.
In Fig. \ref{thin_torus}~(c), we list the Chern numbers for several $\phi=P/Q$ 
computed numerically 
using the method of \cite{Fukui_Hatsugai, Niu_Thouless_Wu}.

We next show that the spatially localized state along $(1,-1)$ direction can be constructed from the solution of Eq. (\ref{midband_eq}). 
Due to the fact $q^Q-q^{-Q}=0$, we can take $\{u_l \}^{2Q-1}_{l=0}$ to be of the form: $u_l=v_l \in \mathbb{R}$ if $0 \le l \le Q-1$ and zero otherwise. 
This solution is degenerate with the other one: $u_l=(-1)^l v_{l-Q}$ if $Q \le l \le 2Q-1$ and zero otherwise~\cite{P_Q_rest}.
The vector $\{v_l\}^{Q-1}_{l=0}$ is normalized as $\sum^{Q-1}_{l=0}v^2_l=1$. 
From those solutions, a localized Wannier state extending from $m-n=j$ to $j+Q-1$ can be constructed as
\hal{changed ``the Wannier state'' into `` a Wannier state''.
The following equation is not quite logical, since it does not define $d$.
But I leave it as it is since I believe no one would be confused.}
\begin{equation}
\hspace{-0.71mm}d^\dagger_{j,\sigma}|\Phi_0\ra=\frac{1}{\sqrt{Q}}\sum^{Q-1}_{l=0}\sum^{2Q-1}_{k=0} \chi_{j+l,k}\, (iq^l)^k v_l c^\dagger_{[j+l,k],\sigma}|\Phi_0\ra,
\end{equation}
where $c^\dagger_{[i,k],\sigma}\equiv c^\dagger_{(\frac{k+i}{2},\frac{k-i}{2}),\sigma}$ and $\chi_{i,k}=1$ if $i$ and $k$ have the same parity and $0$ otherwise. One can easily show that $\{d_{j,\sigma}, d^\dagger_{l,\sigma'}\}=\delta_{jl}\delta_{\sigma \sigma'}$ ($0 \le j, l \le 2L-1$). 
From the Perron-Frobenius theorem, one also finds that the lowest eigenvalue for Eq. (\ref{midband_eq}) is two-fold degenerate. 
This implies that the lowest energy of $H_{\rm hop}$ is $2L$-fold degenerate. Therefore, $d$-states form a complete basis for the lowest flat bands. 


Let us now study the effect of electron correlation within the Hubbard model. 
We define the Hubbard Hamiltonian $H$ by 
\begin{equation}
H=H_{\rm hop}+U\sum_{m,n}n_{(m,n),\ua}n_{(m,n),\da}
\end{equation}
with $U>0$. 
If the total number of electrons $N_{\rm e}$ is $2L$, the ferromagnetic state constructed from the lowest flat band of $H_{\rm hop}$, 
$|\Psi\ra=\prod^{2L-1}_{j=0}d^\dagger_{j,\ua}|\Phi_0\ra$, 
is a ground state of $H$. 
To go further and show that all the ground states are ferromagnetic, we
make use of  theorem due to Mielke \cite{Mielke2}. 
\hal{changed}
The theorem asserts that if the single-particle density matrix constructed from $|\Psi\ra$ as 
\begin{equation}
\rho_{\mathbf{r},\mathbf{r'}}=\frac{1}{N_{\rm e}}\la \Psi|c^\dagger_{(m,n),\ua}c_{(m',n'),\ua}|\Psi \ra
\end{equation}
is irreducible, $|\Psi\ra$ is the unique ground state of $H$ with $N_{\rm e}=2L$ up to the  trivial degeneracy from the SU(2) symmetry~\cite{Mielke_proof}. 
Here, $\mathbf{r}$ and $\mathbf{r'}$ denote $(m,n)$ and $(m',n')$, respectively. 
The diagonal matrix element, the electron density, is uniform and obtained as $\rho_{\mathbf{r},\mathbf{r}}=1/(2QL)$. 
To show that $\rho_{\mathbf{r},\mathbf{r'}}$ is irreducible, it is sufficient to show that any 
matrix element corresponding to a pair of nearest neighbor sites is nonzero. 
Moreover, $\rho_{\mathbf{r},\mathbf{r'}}$ is Hermitian, we have only to study the case of $m'-n'=m-n+1$, $m'+n'=m+n \pm1$. 
In this case, an explicit calculation gives 
\begin{equation}
\rho_{\mathbf{r},\mathbf{r'}}=\frac{q^{m+n}}{2QL} \sum^{Q-2}_{l=0}i^{\pm 1}q^{\pm(l+1)}v_l v_{l+1}.
\end{equation} 
Recalling that $\{ v_l \}^{Q-1}_{l=0}$ satisfy Eq. (\ref{midband_eq}), 
we find ${\rm Re}[\sum^{Q-2}_{l=0}i^{\pm 1}q^{\pm (l+1)}v_l v_{l+1}]=\epsilon_1/4$, where $\epsilon_1$ is the single-particle energy of the lowest flat band. 
For even $Q$, $\epsilon_1 \ne 0$ can be shown by deriving the characteristic equation for $\epsilon(p)$ from Eq. (\ref{midband_eq}) and hence $\rho_{\mathbf{r},\mathbf{r'}}\ne 0$. 
Therefore, via Mielke's theorem, the ferromagnetic state $|\Psi\ra$ is the unique ground state of $H$ up to the spin degeneracy. 
This ground state corresponds to the quantum Hall ferromagnet for the filling factor $\nu=1/P$ with odd $P$. 
So far, we have studied the square lattice model on the thin torus. 
However, our argument can be generalized to other lattice geometry such as the honeycomb lattice which is relevant to the quantum Hall ferromagnetism in graphene~\cite{Nomura_MacDonald}. 
\section{Conclusion}
To conclude, we have studied two classes of Hubbard models with a lowest flat band separated from the other bands by nonvanishing gap originating from the magnetic flux or spin-orbit coupling. In the first class, 
we have shown a systematic way to construct tight binding models with gapped flat bands using line graphs. We have proved that the Chern number of the flat bands is zero in this class of tight binding models. 
We have also studied the effect of the on-site Coulomb interaction 
for two particular cases: i) Kagom\'e ladder and ii) Two-dimensional checkerboard lattice, 
and have rigorously shown that the ground states are ferromagnetic when the lowest flat band is half-filled. 

In the second class, we found the construction of the tight binding models 
embedded on a thin torus, in which all the bands are flat. 
Each flat-band manifold is spanned by the states localized in one direction while delocalized in the other. 
This is reminiscent to the LLL wave functions on a torus. 
We have numerically calculated the Chern number of the lowest band 
and found that it can be nontrivial. 
The lowest flat bands also allow us to study the effect of the on-site Coulomb interaction nonperturbatively. Applying the theorem of Mielke, we have 
shown that the ground states are ferromagnetic when the lowest flat band is half-filled. 
Although our model only reproduces integer quantum Hall systems, it would be  interesting to explore lattice realizations of fractional quantum Hall systems where the next nearest neighbor interaction is probably important~\cite{Assaad}.

\acknowledgments
The authors are grateful to A.~Mielke, N.~Nagaosa, and K.~Nomura   
for their valuable comments and discussions. 
This work is supported in part by Grant-in-Aids (No. 20740214)
from the Ministry of Education, Culture, Sports, Science and Technology of Japan.  HK is supported by the JSPS Postdoctral Fellow for Research Abroad.


\begin{thebibliography}{99}
\bibitem{twofold_graph}
A graph $G$ is twofold connected if and only if one cannot divide $G$ into
disconnected graphs by removing a single vertex.
\bibitem{properties_of_T}
By $|S|$ we denote the number of elements in  a set $S$.
\bibitem{cal_of_Tvv'}
\hal{This footnote is not really necessary, but it is kind for the readers
to say something like this.}
As for the TBM on $G$, one easily finds $(\sfT)_{vv'}=\exp[2\pi i\phi_{vv'}]$
when $\la vv' \ra\in E$, which shows that there is a constant flux.
The calculation of $\sfTL$ is similar but slightly complicated.
\bibitem{proof_of_1st_ineq}
Let $(\varphi_v)_{v \in V}$ be a vector on $V$.
Then, the expectation value of $\sfT(x)-\sfT(0)$ for $(\varphi_v)$ is
$\sum_{e=\langle vv'\rangle \in E_2} |B^\ast_{ve}\varphi_v+B^\ast_{v'e}\varphi_{v'}|^2$, from which the first inequality follows.
\bibitem{P_Q_rest}
Note that $P$ is odd because $P$ and $Q$ are mutually prime.
\bibitem{Mielke_proof}
While the original theorem has been proved for the Hubbard model with real hoppings, it can be generalized to the case of complex hopping relevant to our problem.
\end{thebibliography}
\end{document}